\documentclass[twocolumn,prl,showpacs,amsmath,amssymb,floatfix]{revtex4}
\usepackage{graphicx}

\begin{document}

\title{Resonant Nucleation}

\author{Marcelo Gleiser}
\email{gleiser@dartmouth.edu}

\affiliation{Department of Physics and Astronomy, Dartmouth College,
Hanover, NH 03755, USA}

\author{Rafael C. Howell}
\email{rhowell@lanl.gov}

\affiliation{Materials Science and Technology Division, Los Alamos National Laboratory, 
Los Alamos, NM 87545, USA}

\date{\today}

\begin{abstract}
We investigate the role played by fast quenching on the decay of metastable (or false vacuum) 
states. Instead of the exponentially-slow decay rate per unit volume,
$\Gamma_{\rm HN} \sim \exp[-E_b/k_BT]$ ($E_b$ is the free energy of the critical bubble), predicted
by Homogeneous Nucleation theory, we show that under fast enough quenching the decay rate is a 
power law $\Gamma_{\rm RN} \sim \left[E_b/k_BT\right]^{-B}$, where $B$ is weakly sensitive to the 
temperature. For a range of parameters, large-amplitude oscillations about the metastable state
trigger the resonant emergence of coherent subcritical configurations. Decay mechanisms for 
different $E_b$ are proposed and illustrated in a (2+1)-dimensional scalar field model. 
\end{abstract}


\maketitle

\noindent
{\bf Introduction.}
Few topics in physics have the range of applicability of first order (or discontinuous)
phase transitions. From materials science to particle physics and
cosmology, the fact that a large number
of physical systems can be described as having two phases separated by an energy barrier
has been an active topic of research for decades \cite{gunton} \cite{liquids} \cite{cosmopt}.
Much of the
theoretical work in this field is derived in one way or another from the theory of 
Homogeneous Nucleation (HN) \cite{gunton} \cite{langer}. 
HN assumes that the system is initially localized in a
spatially-homogeneous metastable state,
that is, that only small fluctuations about the local equilibrium state, $\phi_m$, exist: 
one computes the partition function $Z=-T\ln {\cal F}$
summing only over quadratic fluctuations about $\phi_m$. ${\cal F}$ is the free energy,
given by the path integral $\int {\cal D}\phi \exp[-E[\phi]/T]$, where $E[\phi]$ is the
free energy functional of configuration $\phi$. [$k_B =c=\hbar=1$ throughout.]
In relativistic field theory,
false vacuum decay has been examined both at zero
\cite{coleman} and finite temperature \cite{linde} by a large number of authors 
\cite{decay1}.
In general,
the HN approximation is adopted from the start. At finite
temperatures, one uses the
well-known exponential decay rate per unit volume,
$\Gamma_{\rm HN} \simeq T^{(d+1)}\exp[-E_b/T]$,
where $E_b$ is the free energy barrier for the decay, or the energy of the critical
bubble or bounce, $\phi_b(r)$, the solution to the equation
\begin{equation}
\phi''+\frac{d-1}{r}\phi' = \frac{\partial V[\phi]}{\partial\phi}~,
\end{equation}
\noindent
with appropriate boundary conditions.
A prime denotes derivative with respect to the 
$d$-dimensional radial coordinate (sphericity is energetically
favored) and $V[\phi]$ is the effective potential that sums
over thermal and quantum contributions when applicable. At $T=0$, one has
a purely quantum vacuum decay, and the pre-factor $T^{(d+1)}$ is roughly approximated by
$M^{(d+1)}$, the relevant mass scale, while $E_b/T$ is substituted by $S_E[\phi_b]$,
the $(d+1)$-dimensional
Euclidean action of the bounce configuration.

In the present work we examine what happens if one relaxes
the HN approximation that the initial state is well-localized
about equilibrium. We subject the system to an instantaneous quench, 
equivalent to a sudden
change of potential from a single to an asymmetric double well. This should
be contrasted with the work of ref. \cite{Felder-Linde} which studies quenches
in models {\it without} a barrier separating symmetric and broken-symmetric
states, and thus with spinodal decomposition dynamics. Although
in this first study we will only consider instantaneous quenches, we expect
our results to carry on at least partially to slower quenches, 
so long as the quenching rate
$\tau_{\rm quench}$ is faster than the relaxation rate of the field's zero mode,
$\tau_0$. Why the field's zero mode? With the longest wavelength it is the 
slowest to equilibrate: as $\tau_{quench}\rightarrow \tau_0$, the
system will remain in equilibrium. For 
appropriate choices of parameters, the rapid
quench will induce large-amplitude oscillations of the field's 
zero mode \cite{gleiser-howell}. 
Due to the nonlinear potential, energy will be
transferred from the zero mode to higher $k$-modes. As observed in reference
\cite{gleiser-howell}, this transfer of
energy results in the synchronous emergence of oscillon-like
configurations \cite{oscillons}. [We urge the reader to consult reference
\cite{gleiser-howell} for details.] For small enough double-well asymmetry,
these localized field configurations act as precursors
for the nucleation of a critical bubble, greatly reducing the decay time-scale. In the 
simulations we examined, the critical bubble emerges as two or more subcritical oscillons
coalesce, or, for larger asymmetries, as a single oscillon becomes critically
unstable to growth. If the asymmetry is too large, the field crosses directly to 
the global minimum.

\noindent
{\bf The Model.}
Consider a (2+1)-dimensional real scalar field (or scalar order parameter)
$\phi({\bf x},t)$ evolving under the influence of a potential 
$V(\phi)$. The continuum Hamiltonian is conserved and the total energy of a given field
configuration $\phi({\bf x},t)$ is,
\begin{equation}
H[\phi]=\int d\,^2x\left[\frac{1}{2}(\partial_{t}\phi)^2+\frac{1}{2}(\nabla\phi)^2+V(\phi)\right],
\label{H}
\end{equation}
where $V(\phi)=\frac{m^2}{2}\phi^2-\frac{\alpha}{3}\phi^3+\frac{\lambda}{8}\phi^4$ 
is the potential energy density.
The parameters $m$, $\alpha$, and $\lambda$ are positive definite and temperature independent. 
It is helpful to introduce the dimensionless variables $\phi'=\phi\,\sqrt\lambda/m$,
$x'=xm$, $t'=tm$, and $\alpha'=\alpha/(m\sqrt\lambda)$ (We will henceforth drop the primes). 
Prior to the quech, $\alpha=0$ and the potential is an anharmonic single
well symmetric about $\phi=0$. The field is in thermal equilibrium
with a temperature $T$.  At the temperatures considered, the fluctuations of the field are
well approximated by a gaussian distribution, with $\langle\phi^2\rangle = aT$ ($a=0.51$ and
can be computed numerically \cite{gleiser-howell_long}).  As such, within 
the context of the Hartree approximation \cite{aarts}, the momentum and field modes in $k$-space 
can be obtained from a harmonic effective potential, and 
satisfy $\langle|{\bar \pi}(k)|^2\rangle =T$ and 
$\langle |\bar{\phi}({\bf k})|^2\rangle=\frac{T}{k^2+m_H^2}$, respectively. 
The Hartree mass $m_H^2=1+\frac{3}{2}\langle\phi^2\rangle$ depends on the 
magnitude of the fluctuations (and thus $T$).  Hereafter we will refer
to a particular system by its initial temperature. All results are ensemble averages over 100
simulations.

If $\alpha \neq 0$, the
${\cal Z}_2$ symmetry is explicitly broken. 
When $\alpha =1.5\equiv \alpha_c$, the potential is a symmetric double-well (SDW), with two 
degenerate minima. 
The quench is implemented at $t=0$ by setting $\alpha > \alpha_c$, whereby the potential is 
asymmetric (ADW) about the barrier separating the two minima.  The Hartree approximation gives an 
accurate description of
the evolution of the area-averaged field $\phi_{\rm ave}(t)$ and its fluctuations for early times 
after 
the quench, in which the distribution of
fluctuations remains gaussian and the dynamics are governed by an effective potential,
\begin{eqnarray}
\label{Veff}
V_{\rm eff}\left(\phi_{\rm ave},m_{\rm H}^2\right)&=& 
\left[1-m_{\rm H}^2(t)\right]\phi_{\rm ave}
+\frac{1}{2}\,m_{\rm H}^2(t)\,\phi_{\rm ave}^2 \nonumber \\
&-&\frac{\alpha}{3}\,\phi_{\rm ave}^3+\frac{1}{8}\,\phi_{\rm ave}^4~.
\end{eqnarray}
The quench shifts the local minimum from $\phi_{\rm ave}=0$ to a positive value, 
but also introduces a new global minumum to the system. [See inset in Fig. \ref{decay-phi}.]

\begin{figure}
\includegraphics[width=3in]{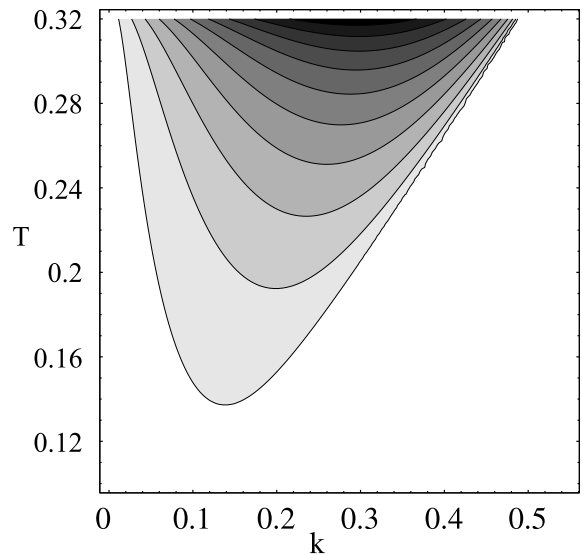}
\caption
{Lines of constant amplification rate $\eta$
for small-amplitude modes at various temperatures,
beginning with $\eta_{\rm min}=2.8\times 10^{-2}$ for the bottom-most contour and increasing in
increments of $\Delta\eta=1.3\times 10^{-2}$.}
\label{parares}
\end{figure}

\begin{figure}
\includegraphics[width=3in]{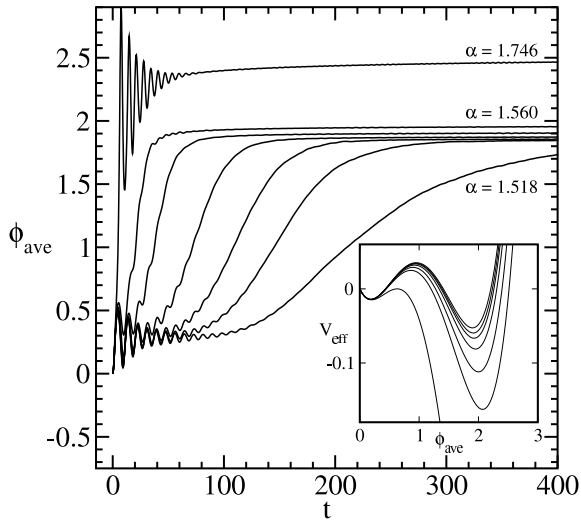}
\caption
{The evolution of the order parameter $\phi_{\rm ave}(t)$ 
at $T=0.22$ for several values of the asymmetry. 
From left to right, $\alpha= 1.746, 1.56, 1.542, 1.53,
1.524, 1.521, 1.518$. The inset shows $V_{\rm eff}$ for the same values.
}
\label{decay-phi}
\end{figure}

\noindent
{\bf Nucleation of Oscillons.}
The quench induces oscillations 
in $\phi_{\rm ave}$ about the new local minimum, which eventually dampen due to nonlinear
scattering with higher $k$-modes.
At early times small fluctuations satisfy a Mathieu equation in $k$-space
\begin{equation}
\label{flucteq}
\ddot{\delta\phi}=-\left [k^2+
V_{\rm eff}''\left[\phi_{\rm ave}(t)\right]\right]\delta\phi~,
\end{equation}
and, depending on the wave number and parametric oscillations of $\phi_{\rm ave}(t)$, can undergo
exponential amplification ($\sim e^{\eta t}$).
In Fig. \ref{parares} we show the lines of constant amplification rate 
for different wave numbers and
temperatures when $\alpha=\alpha_c$ and $V_{\rm eff}$ is defined by the initial thermal
distribution.  At low temperatures $T\lesssim 0.13$, no modes are ever amplified. As 
the temperature
is increased, so is the amplitude and period of oscillation in $\phi_{\rm ave}$, gradually
causing the band $0<k<0.48$ to resonate and grow.  A full description of the coupled dynamics
of $\phi_{\rm ave}(t)$ and $\delta\phi({\bf x},t)$ when $\alpha=\alpha_c$ 
is given in ref. \cite{gleiser-howell_long}.
Furthermore, for large enough
temperatures ($T\gtrsim 0.13$) large-amplitude fluctuations about the
zero mode probe into unstable regions where $V_{\rm eff}''<0$, which also promote
their growth.  Note that this is very distinct from spinodal decomposition,
where competing domains of the metastable and stable
phases coarsen \cite{gunton}. Instead, for the values of $T$
and $\alpha$ considered, $\phi_{\rm ave}$ continues to oscillate about the metastable minimum until
a critical bubble of the stable phase grows to complete the transition.

These two processes result in the synchronous emergence
of oscillon-like configurations (Fig. 3 in ref. \cite{gleiser-howell}), 
long-lived time-dependent localized field
configurations which are well-described by gaussian profiles, 
$\phi_{\rm osc}(t,r)\simeq \phi_a(t)\exp[-r^2/R^2]$ \cite{oscillons}.
To strengthen our argument, note that within this gaussian ansatz,
an oscillon is comprised by modes within the band $0< k\leq 2/R$.
One of us has recently shown that, in $d$ dimensions and for a potential $V$,
the radius of an oscillon satisfies 
$R^2 \geq d/[\frac{1}{2}(2^{3/2}/3)^d(V''')^2/V^{IV}-V'']$ \cite{d_oscil}.
For the potential of eq. \ref{Veff} and $d=2$, we obtain that the 
related band of wave numbers is,
$0<k\lesssim 0.66$. Referring back to Fig. \ref{parares}, the reader can verify 
that these are also approximately
the modes excited by parametric resonance. 

\noindent
{\bf Resonant Nucleation.}
Having established that oscillons emerge after the quench,
we can examine their role as precursors of metastable decay. Unless the potential has a
large asymmetry, oscillons are typically sub-critical fluctuations; as will
be discussed below, a critical nucleus
may appear only due to the coalescence of two or more oscillons.
It should be clear, however, that their appearance renders the homogeneity 
assumption of HN theory inapplicable:
the metastable background is far from homogeneous and the critical energy barrier must be
renormalized \cite{gleiser-heckler}. In other words, {\it a rapid quench or cooling leads to
departures from the usual HN assumptions.} As we describe next, 
the decay rate of the quenched system may
be much faster than what is predicted by HN theory.

In Fig. \ref{decay-phi} we show the evolution
of the order parameter $\phi_{\rm ave}(t)$ 
as a function of time for several values of asymmetry, $1.518 \leq \alpha \leq 1.746$, for $T=0.22$.
Not surprisingly,
as $\alpha\rightarrow \alpha_c=1.5$, the field remains longer in the metastable state, since the 
nucleation energy
barrier $E_b\rightarrow \infty$ at $\alpha_c$. However, 
a quick glance at the time axis shows the fast decay time-scale, of order $10^{1-2}$. 
For comparison, for $1.518\leq \alpha \leq 1.56$, homogeneous nucleation would predict 
nucleation time-scales of order $\sim 10^{28} \geq \tau_{\rm HN}\sim \exp[E_b/T]\geq 10^{12}$ 
(in dimensionless units). [The related
nucleation barriers with the effective potential
are $E_b(\alpha=1.518)=14.10$ and $E_b(\alpha=1.56)=5.74$.]
While for smaller asymmetries $\phi_{\rm ave}(t)$ displays similar oscillatory behavior to the 
SDW case before transitioning to the global minimum, as
$\alpha$ is increased the number of oscillations decreases. 
For large asymmetries, $\alpha\geq 1.746$, the entire field
crosses over to the global minimum without any nucleation event, resulting in oscillations about
the global minimum.  The inset in Fig. \ref{decay-phi} shows that the barrier is just low enough 
for this to occur.

\begin{figure}
\includegraphics[width=3in]{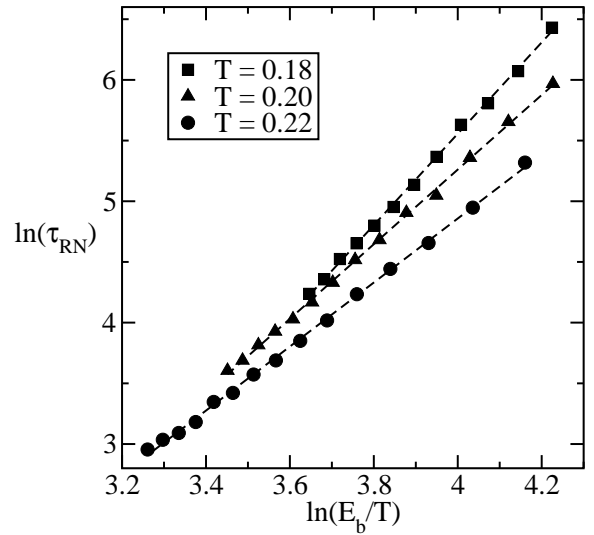}
\caption
{Decay time-scale $\tau_{\rm RN}$ as a function of critical nucleation 
effective free-energy barrier $E_b/T$ at $T=0.18$, $0.20$, and $T=0.22$. 
The best fits (dashed lines) are power-laws with exponents $B\simeq 3.762$, $3.074$, and $2.637$,
respectively.}
\label{powerlaw}
\end{figure}

In Fig. \ref{powerlaw} we show the ensemble-averaged nucleation time-scales 
for resonant nucleation, $\tau_{\rm RN}$, as a function
of the nucleation barrier (computed with eq. \ref{Veff}), 
$E_b/T$, for the temperatures
$T=0.18$, $0.20$, 
and $0.22$. [For temperatures above $T=0.26$ one is in the vicinity of the
critical point in which no barrier exists.] The nucleation time was measured when $\phi_{\rm ave}$ 
crosses the
maximum of $V_{\rm eff}$.
The best fit is a power law, 
$\tau_{RN} \propto
(E_b/T)^B$, with $B=3.762 \pm 0.016$ for $T=0.18$, $B=3.074 \pm 0.015$ for $T=0.20$, 
and $B=2.637 \pm 0.018$ for $T=0.22$. This simple power
law holds for the same range of temperatures where we have observed the synchronous emergence 
of oscillons. It is not surprising that the exponent $B$ increases with decreasing $T$, 
since the synchronous emergence of oscillons becomes less pronounced and eventually vanishes.
In these cases
we should expect a smooth transition into the exponential time-scales of HN.
We present below what we believe is the mechanism
by which the transition completes for different nucleation barriers.

\begin{figure}
\includegraphics[width=2.5in]{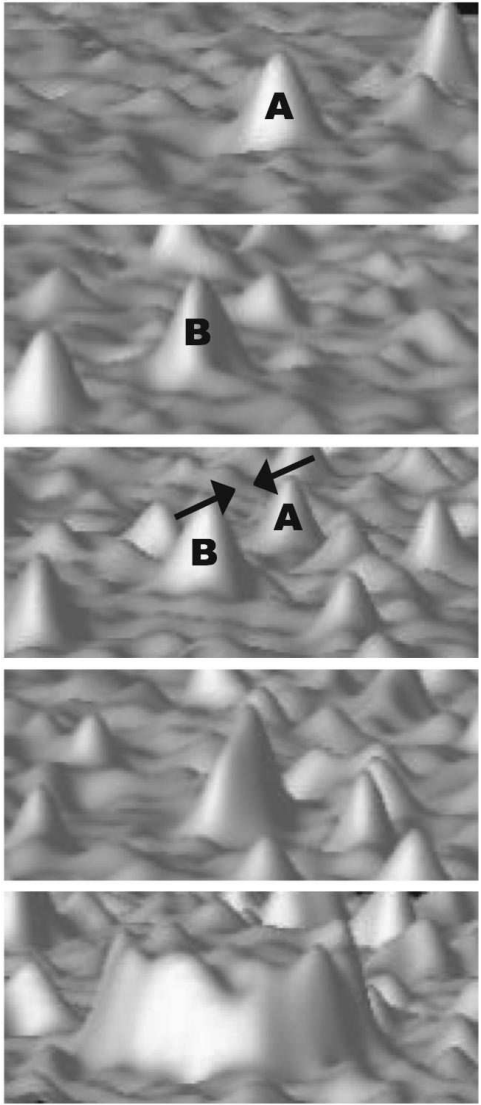}
\caption
{Two oscillons coalesce to form a critical bubble. 
First two frames from top show oscillons A and B. Third and fourth
frames shows A and B coalescing into a critical bubble. 
Final frame shows growth of bubble expanding into metastable state.}
\label{oscil_coal}
\end{figure}

First, for $\alpha \rightarrow \alpha_c$, 
the radius of the
nucleation bubble diverges, $R_b \rightarrow \infty$. When fast quenching induces
large-amplitude fluctuations of the field's zero mode,
the system doesn't approach the global minimum
through a random search in configuration space as is the case in HN. Instead,
we argue that oscillons will induce the nucleation of a critical fluctuation. 
The way in which 
this happens depends on the magnitude of the nucleation barrier: 
for nearly degenerate potentials, $\alpha_c <\alpha \lesssim\alpha_{\rm I}$,
the critical nucleus has a much larger radius than a typical oscillon;
it will appear as two or more oscillons coalesce. We call this Region I,
defined for $R_b \geq 2R_{\rm osc}$, where $R_{\rm osc}$ is the minimum
oscillon radius computed from ref. \cite{d_oscil}.
Fig. \ref{oscil_coal} illustrates this mechanism.
Two oscillons, labeled A and B, join to become a critical nucleus.
They diffuse through the lattice and form bound states, 
somewhat as in kink-antikink breathers in 1d field
theory \cite{campbell}. [The interested reader can see simulation movies at 
http://www.dartmouth.edu/$\sim$cosmos/oscillons.]
We are currently attempting to estimate the diffusion 
and coalescence rate of
oscillons on the lattice so that we can compute the power law decay rate 
analytically.

\begin{figure}
\includegraphics[width=3in]{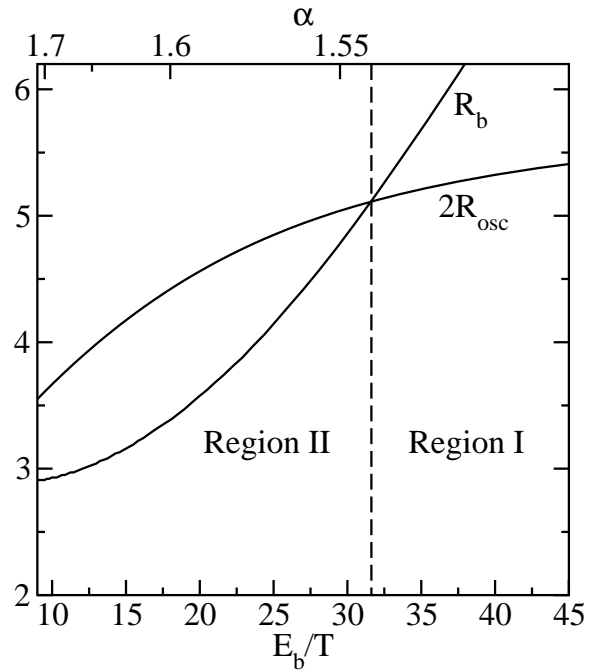}
\caption
{Radius of critical bubble ($R_b$) and twice the minimum oscillon radius ($2R_{\rm osc}$)
as a function of its energy barrier and related values
of $\alpha$ at $T=0.22$. For $\alpha \gtrsim 1.547$ one cannot easily distinguish between 
an oscillon and a critical bubble.}
\label{radius}
\end{figure}

As $\alpha$ is increased further,
the radius of the critical nucleus decreases, approaching that of an oscillon. 
In this case, a single
oscillon grows unstable to
become the critical nucleus and promote the fast decay of the metastable state:
there is no coalescence. We call this Region II, $\alpha_{\rm I}<\alpha\lesssim\alpha_{\rm II}$,
$R_b < 2R_{\rm osc}$.
This explains
the small number of oscillations on $\phi_{\rm ave}(t)$ as $\alpha$ is increased 
[cf. Fig. \ref{decay-phi}].
To corroborate our argument, in Fig. \ref{radius} we contrast the critical 
nucleation radius with that
of oscillons as obtained in ref. \cite{d_oscil}, 
for different values of effective 
energy barrier and related values of $\alpha$ at $T=0.22$. 
The critical nucleus radius $R_b$ is
equal to $2R_{\rm osc}$ for $\alpha = 1.547$. This defines
the boundary between Regions I and II:
for $\alpha \gtrsim \alpha_{\rm I}$
a single oscillon may grow into a critical bubble. Finally,
for $\alpha \gtrsim \alpha_{\rm II} = 1.746$ the field crosses over to the global
minimum without any nucleation event.

How will the efficiency of the mechanism decrease as $\tau_{\rm quench}
\rightarrow \tau_0$? What happens when the quench is induced by cooling as,
for example, in the early universe? Does the power law behavior obtained here 
still hold for $d=3$? We intend
to address these and related questions in the near future.

MG was partially supported by a NSF grant PHYS-0099543. We would like to thank 
Jim Krumhansl, friend and mentor, for his insightful remarks.


\begin{thebibliography}{99}

\bibitem{gunton} J. D. Gunton, M. San Miguel, and P. S. Sahni, in {\it Phase Transitions
and Critical Phenomena}, Ed. C. Domb and J. L. Lebowitz, v. 8 (Academic Press, London, 1983); J. D.
Gunton, J. Stat. Phys. {\bf 95}, 903 (1999); J. S. Langer, in {\it Solids Far from Equilibrium}, Ed. C.
Godr\`eche (Cambridge University Press, Cambridge, 1992).

\bibitem{liquids} P. G. de Gennes, {\it The Physics of Liquid Crystals}
(Oxford University Press, Oxford, 1993).

\bibitem{cosmopt} A. Vilenkin and E. P. S. Shellard, {\it Cosmic Strings
and Other Topological Defects} (Cambridge University Press, Cambridge, 1994).

\bibitem{langer} J. S. Langer, Ann. Phys. (NY) {\bf 41}, 108 (1967);
{\it ibid.} {\bf 54}, 258 (1969).

\bibitem{coleman} S. Coleman,  Phys. Rev. D{\bf 15}, 2929 (1977);
C. Callan and S. Coleman, Phys. Rev. D{\bf 16}, 1762 (1977).

\bibitem{linde} A. Linde, Nucl. Phys. B{\bf 216}, 421 (1983);
[Erratum: B{\bf
223}, 544 (1983)].

\bibitem{decay1} Here is an incomplete list of references:
M. Stone,
Phys. Rev. D {\bf 14} (1976) 3568;
Phys. Lett. B {\bf 67} (1977) 186;
M. Alford and M. Gleiser, Phys. Rev. 
D {\bf 48}, 2838 (1993); J. Baacke and V. Kiselev, Phys. Rev. D {\bf 48}, 5648 (1993);
H. Kleinert and I. Mustapic,
Int. J. Mod. Phys. {\bf A11} (1996) 4383,
and references therein;  A. Strumia and N. Tetradis,
Nucl. Phys. B {\bf 554}, 697 (1999); {\it ibid.} {\bf 560}, 482 (2000);
Sz. Bors\'anyi et al., Phys. Rev. D {62}, 085013-1 (2000).
Y. Bergner and Luis M. A. Bettencourt,
Phys. Rev. D {\bf 69} (2004) 045012.

\bibitem{Felder-Linde} G. Felder {\it et al.}
Phys. Rev. Lett. {\bf 87}, 011601 (2001).

\bibitem{gleiser-howell} M. Gleiser and R. Howell, Phys. Rev. E {\bf 68}, 065203(R) (2003).

\bibitem{oscillons} M. Gleiser, Phys. Rev. D {\bf 49}, 2978 (1994); 
E. J. Copeland, M. Gleiser, and H. R. Muller, Phys. Rev. D {\bf 52}, 1920 (1995);
E. B. Bogomol'nyi, Sov. J. Nucl. Phys. {\bf 24}, 449 (1976).

\bibitem{gleiser-howell_long} M. Gleiser and R. Howell, cond-mat/0310157.

\bibitem{aarts} G. Aarts, G. F. Bonini, and C. Wetterich, Phys. Rev. D {\bf 63}, 025012 (2000);
G. Aarts, G. F. Bonini, and C. Wetterich, Nucl. Phys. B {\bf 587}, 403 (2000).

\bibitem{d_oscil} M. Gleiser, Phys. Lett. B {\bf 600}, 126 (2004).

\bibitem{oscil2d} M. Gleiser and A. Sornborger, Phys. Rev. E {\bf 62}, 1368 (2000);
A. Adib, M. Gleiser, and C. Almeida, Phys. Rev. D {\bf 66}, 085011 (2002).

\bibitem{gleiser-heckler} M. Gleiser and A. Heckler, Phys. Rev. Lett. {\bf 76}, 180 (1996).

\bibitem{campbell} D. K. Campbell, J. F. Schonfeld, and C. A. Wingate, Physica {\bf 9D}, 1 (1983).

\end{thebibliography}
\end{document}